 \documentclass{article}
 \usepackage{spconf}

\usepackage{amsmath,graphicx,hyperref}
\usepackage{booktabs}
\usepackage{multirow}
\usepackage{amssymb}
\usepackage{pifont}
\usepackage{float}
\usepackage{makecell}
\usepackage{hyperref}
\usepackage{cite}


\usepackage{subcaption}

\newcommand{\cmark}{\ding{51}} 
\newcommand{\xmark}{\ding{55}} 

\title{Short-segment speaker verification with \\ pre-trained models and multi-resolution encoder}
\name{Jisoo Myoung, Sangwook Han, Kihyuk Kim, Jong Won Shin\thanks{This work was partly supported by the Institute of Information \& Communications Technology Planning \& Evaluation(IITP)-ITRC(Information Technology Research Center) grant funded by the Korea government(MSIT)(IITP-2026-RS-2021-II211835, 50\%) and IITP grant funded by MSIT (No.2019-0-01842,
Artificial Intelligence Graduate School Program (GIST), 50\%).}}
\address{Gwangju Institute of Science and Technology, Gwangju, Korea}

\begin{document}
\ninept 
%
\maketitle
\begin{abstract}
Speaker verification (SV) utilizing features obtained from models pre-trained via self-supervised learning has recently demonstrated impressive performances.
However, these pre-trained models (PTMs) usually have a temporal resolution of 20 ms, which is lower than typical filterbank features.
It may be problematic especially for short-segment SV with an input segment shorter than 2 s, in which we need to extract as much information as possible from the input with a limited length.
Although there have been approaches to utilize multi-resolution features from the HuBERT models, the window shifts were 20, 40, and 100 ms when the sampling rate was 16 kHz and thus only lower resolution features were considered. 
In this study, we propose an SV system which utilizes PTM features along with filterbank features and those from the multi-resolution time domain encoder with window shifts of 1.56, 3.13, 6.25, and 12.5 ms. 
Experimental results on the VoxCeleb dataset with various input lengths showed consistent improvements over systems with various combinations of input features.
\end{abstract}

\begin{figure*}[!htbp]
    \centering
    \includegraphics[width=\textwidth]{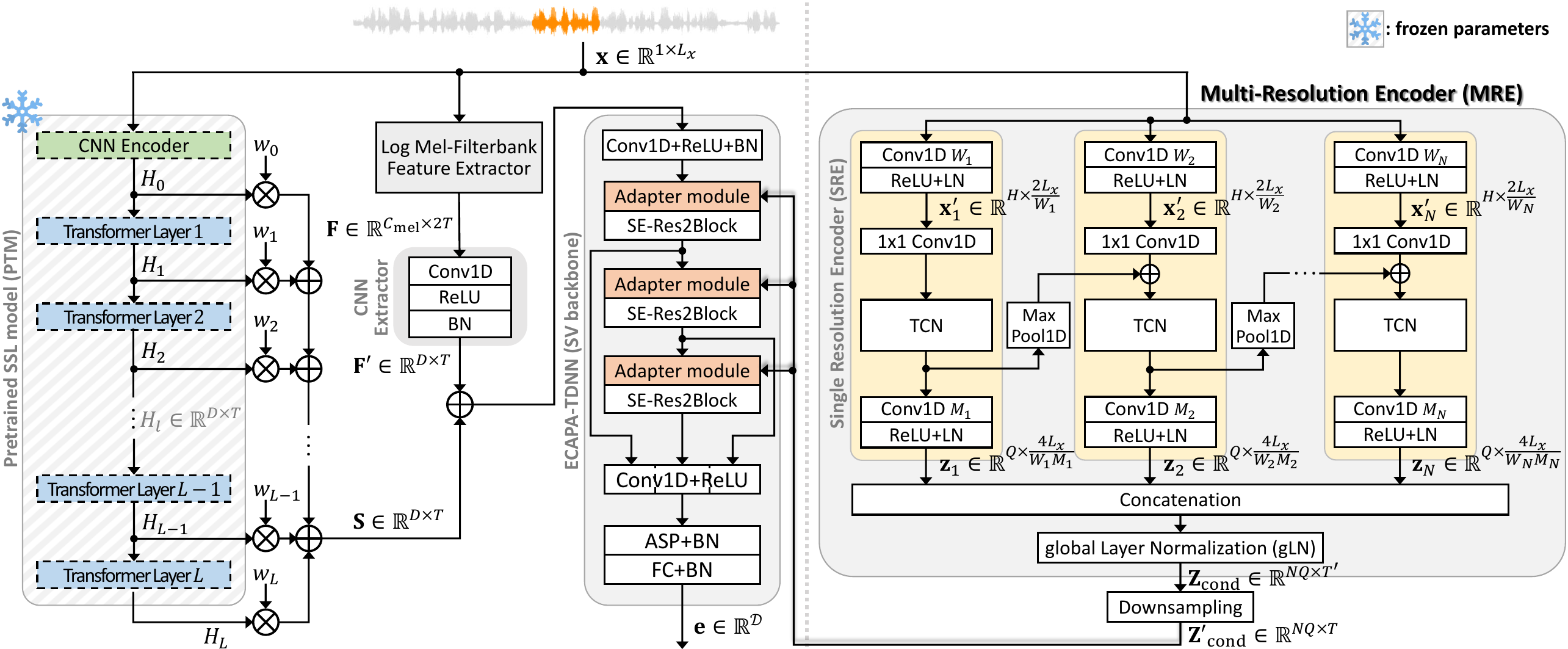}
    \vspace{-0.5cm}
    \caption{Block diagram of the proposed short-segment speaker verification system with a pre-trained model and a multi-resolution encoder. }
    \label{fig:overall_framework}
\end{figure*}
\section{Introduction}
\label{sec:intro}

Speaker verification (SV) is the process of verifying if the speaker for a given input utterance matches the claimed identity 
\cite{sr-ct-sv, ecapa-tdnn, mfa-tdnn, rawnext, ecapa-mre, mr-rawnet, eres2netv2, ssl-sv1, ssl-sv2, ssl-sv3, ssl-sv4, ssl-sv5, ssl-sv6, ssl-sv7, ssl-sv8}.
With the advancement of deep learning, speaker representations can be utilized in terms of auxiliary information for various speech processing tasks such as target speaker extraction \cite{hao2024x}, and speech separation \cite{monaural_ss}.
The performance of the SV heavily depends on the length of the input and enrollment utterances, which makes short-segment SV quite challenging. 
Various approaches have been proposed to extract as much speaker-related information as possible from limited speech signal by handling multi-scale features in temporal or channel directions \cite{mfa-tdnn, rawnext, ecapa-mre, mr-rawnet, eres2netv2}.
In \cite{mfa-tdnn}, multi-scale feature-channel attention (MFA) module is proposed as a front-end module for time delay neural networks (TDNNs) to extract multi-scale information effectively by feeding high-resolution information to lower resolution and emphasizing relevant frequency bins and channels.
RawNeXt \cite{rawnext} produces speaker embeddings enriched with time-spectral information by iteratively and hierarchically aggregating multi-resolution features.
In \cite{ecapa-mre}, the ECAPA-TDNN \cite{ecapa-tdnn} taking log mel-filterbank (FBank) features is modified to incorporate the features extracted by a multi-resolution encoder (MRE) through adapter modules to utilize information obtained with various sizes of windows. 
MR-RawNet \cite{mr-rawnet} improves RawNet3 by replacing its first layer with multi-resolution feature extractor (MRFE), enabling simultaneous consideration of temporal and spectral resolutions.
ERes2NetV2 \cite{eres2netv2} expands and compresses channels to strengthen short-duration feature extraction through a bottleneck-like local feature fusion (BLFF), and then a bottom-up dual-stage feature fusion (BDFF) integrates multiscale feature maps with different receptive fields to capture global information.

Recently, there has been growing interest in SV systems that utilize speech representations extracted from models pre-trained via self-supervised learning (SSL) on large-scale speech data \cite{ssl-sv1, ssl-sv2, ssl-sv3, ssl-sv4, ssl-sv5, ssl-sv6, ssl-sv7, ssl-sv8}.
In particular, various fusion methods to combine layer-wise outputs from pre-trained SSL models (PTMs) to produce the input for the SV models have been proposed \cite{ssl-sv2, ssl-sv3, ssl-sv4, ssl-sv5, ssl-sv6}.
While they exhibited potential of the PTM-based features, the temporal resolutions for these features were fixed to be the same as those for the corresponding PTMs, which is typically 20 ms.
In \cite{hubert-with-mr}, multiple temporal resolutions in HuBERT \cite{HuBERT} have been proposed by pre-training three HuBERT models at different temporal resolutions and fusing their outputs using hierarchical or parallel approaches.
MR-HuBERT \cite{mr-hubert} further extended the original HuBERT with two resolutions through a hierarchical Transformer, where predefined hidden representations were downsampled to incorporate both high- and low-resolution masked unit predictions, respectively.
Although exploiting multiple temporal resolutions for HuBERT was shown to be beneficial for a broad range of speech processing tasks in these papers, the temporal resolutions they utilized corresponded to 20, 40, and 100 ms of frame shifts, which may be too low for the short-segment SV. 
In \cite{ssl-sv4}, the features from the PTMs fused with a fine-grained fusion module (FGFM) were combined with the FBank features with a 10 ms temporal resolution to produce excellent performances.

To utilize the representation capability of the features from the PTMs while exploiting multiple fine temporal resolutions for short-segment SV, we propose to utilize not only the features from PTMs and FBank features but also the features from the MRE.
The weighted summation of the features from the PTMs and the FBank features are used as the input of the ECAPA-TDNN \cite{ecapa-tdnn} as in \cite{ssl-sv4}, while the MRE features are incorporated through adapter modules as in \cite{ecapa-mre}. To focus more on the combination of these three features, we simply combine the layer-wise outputs of the PTMs by weighted summations.
Experimental results on the VoxCeleb dataset showed that the adoption of the FBank features and MRE features were effective to improve the performance of the SV utilizing PTMs for various lengths of the input segments and outperformed previously proposed methods in short-segment SV.
Moreover, an analysis of the learned weights to combine layer-wise outputs of the PTMs for different combinations of the features demonstrated that the MRE and FBank features encouraged PTM features to focus more on the high-level information.

\section{Related Works}
\label{sec:format}

\subsection{Multi-Resolution Encoder (MRE)}
In the short-time Fourier transform (STFT) analysis, a longer window yields a higher spectral resolution but a lower temporal resolution, while a smaller window leads to the opposite trade-off when the overlap ratio between adjacent windows is fixed.
In analogy to the multi-resolution STFT analysis utilizing multiple windows with various sizes, the MRE \cite{ecapa-mre} processes the input speech using $N$ single resolution encoders (SREs), each of which analyzes the input signal using a convolutional layer with a different kernel and hop sizes as illustrated in the right part of Fig. \ref{fig:overall_framework}. 
In the $n$-th SRE, the input speech signal $\mathbf{x}\in \mathbb{R}^{1\times L_x}$ is first encoded with $H$ kernels of size $W_n$ and stride $W_n/2$ to produce $\mathbf{x}_n'\in\mathbb{R}^{H \times \frac{2L_x}{W_n}}$.
The encoded features are converted to $P$ channels through a $1 \times 1$ convolution and subsequently processed by temporal convolutional networks (TCNs) \cite{tcn}.
It is then convolved with $Q$ kernels of size $M_n$ and stride $M_n/2$ to align the temporal dimension by configuring $M_n$ to satisfy $\frac{W_n}{2}\cdot\frac{M_n}{2}=\mathcal S$ where $\mathcal S$ is the frame shift and $T'=L_x/\mathcal S$ is the number of frames. 
The outputs $\textbf{z}_n \in \mathbb{R}^{Q \times T'}$ of the $N$ SREs are concatenated along the channel dimension and then normalized using global layer normalization (gLN) \cite{conv-tasnet} to form a multi-resolution feature  $\textbf{Z}_\text{cond} \in\mathbb{R}^{NQ \times T'}$.

\subsection{Integrated Filterbank and PTM Features}
In \cite{ssl-sv4}, layer-wise speech representations from the PTMs are fused using the FGFM to form a PTM feature $\mathbf{S}$, which is then combined with a 80-dimensional FBank feature $\mathbf{F}$.
$\mathbf{F}$ is extracted with a 25 ms window and a hop size of 10 ms, which makes the number of frames twice compared with those for PTM features. 
To match the temporal and channel dimensions, $\mathbf{F}$ is processed by a CNN extractor, which consists of a 1D convolutional layer, a ReLU activation function, and batch normalization (BN).
The aligned feature $\mathbf{F}'$ and $\mathbf{S}$ are added together with two learnable parameters $\alpha$ and $\beta$ to construct the integrated features, i.e.,
\begin{equation}
\text{Integrated Features}=\alpha \mathbf{S} + \beta \mathbf{F'}.
\end{equation}

\begin{figure}[t]
\centering
\begin{subfigure}[b]{0.48\linewidth}
  \centering
  \includegraphics[width=\linewidth,clip]{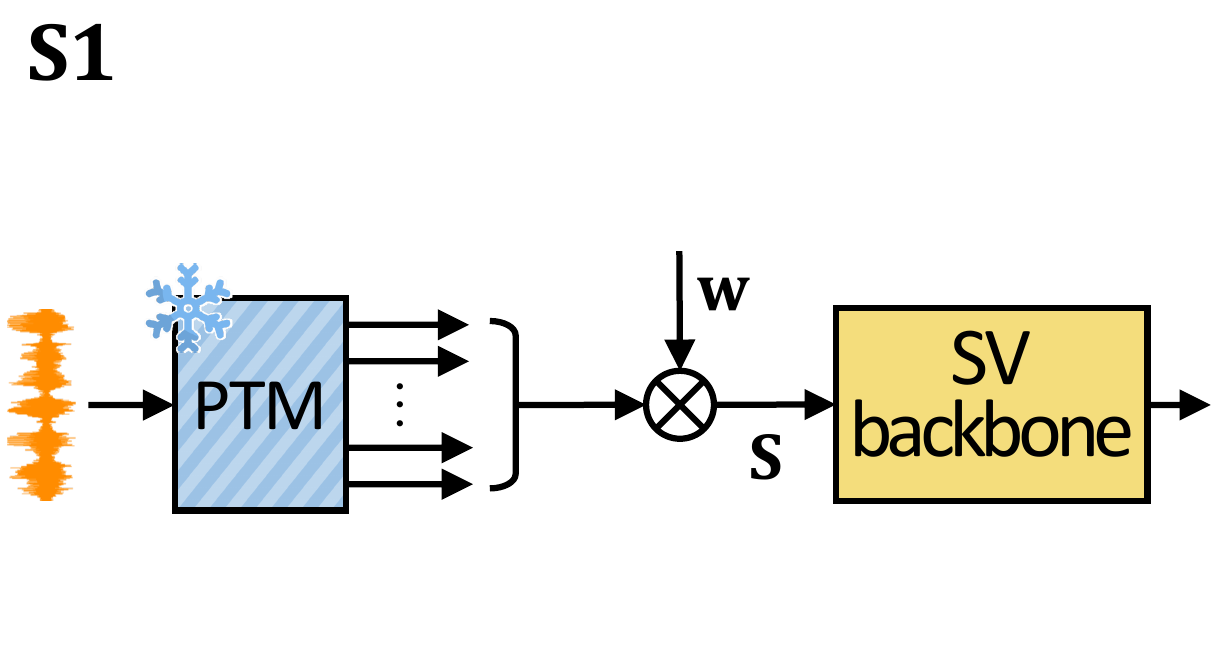}\vspace{0.2 cm}\\[-1.2ex]
  \caption{}
  \label{fig:res_a}
\end{subfigure}\hfill
\begin{subfigure}[b]{0.48\linewidth}
  \centering
  \includegraphics[width=\linewidth,clip]{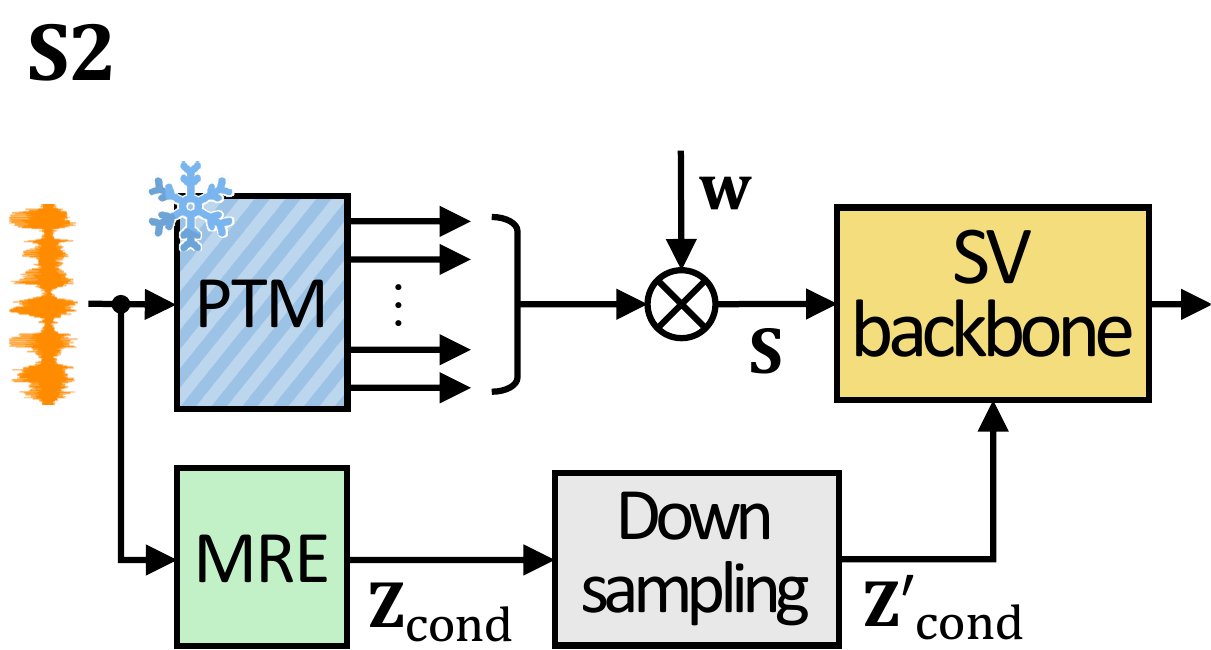}\vspace{0.2 cm}\\[-1.2ex]
  \caption{}
  \label{fig:res_b}
\end{subfigure}
\vspace{0.5cm} 
\begin{subfigure}[b]{0.48\linewidth}
  \centering
  \includegraphics[width=\linewidth,clip]{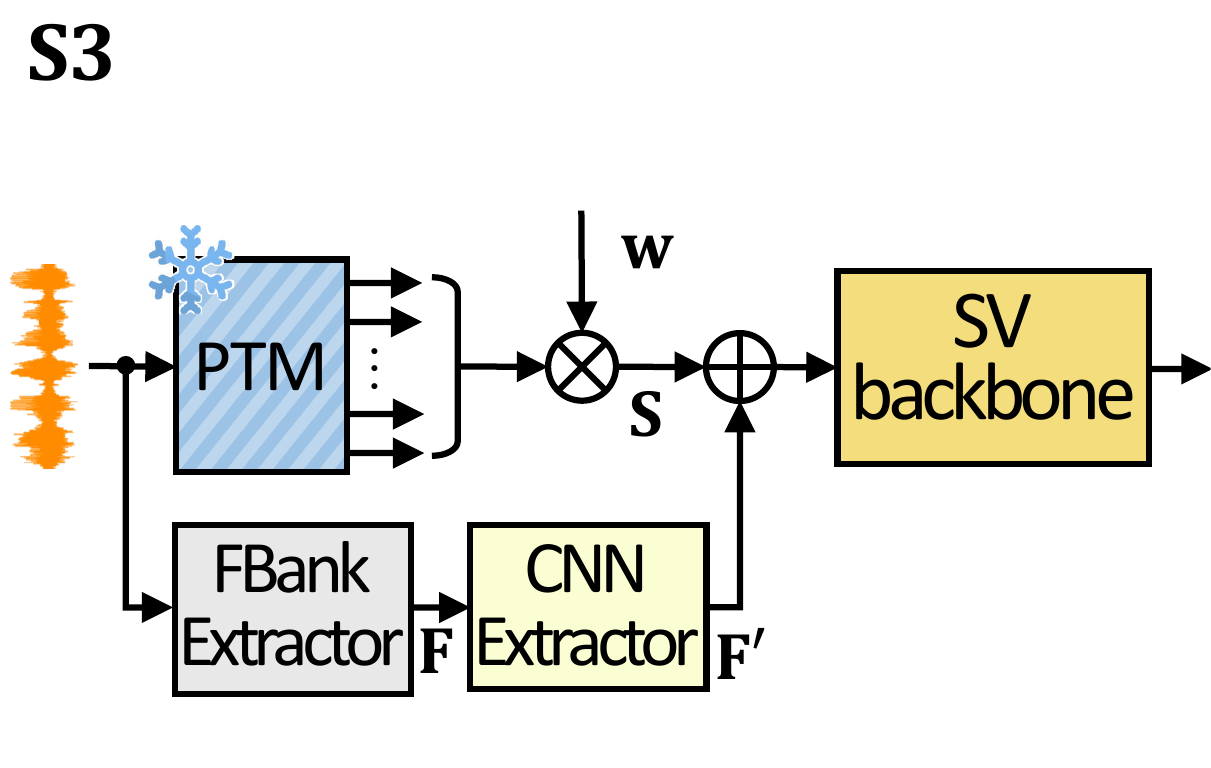}\vspace{0.2 cm}\\[-1.2ex]
  \caption{}
  \label{fig:res_d}
\end{subfigure}\hfill
\begin{subfigure}[b]{0.48\linewidth}
  \centering
  \includegraphics[width=\linewidth,clip]{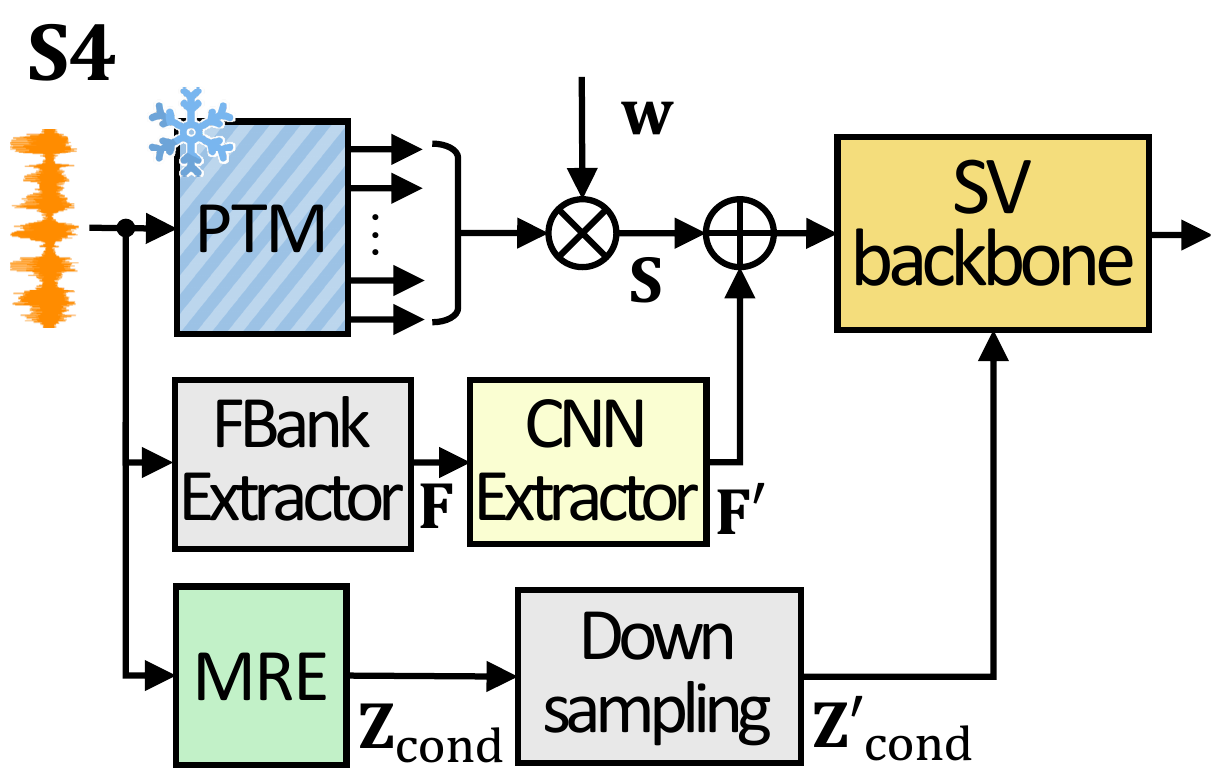}\vspace{0.2cm}\\[-1.2ex]
  \caption{}
  \label{fig:res_d}
\end{subfigure}
\label{fig:s4}
\vspace{-0.7cm}
\caption{Four combinations of input features for speaker verification. }
\label{fig:res}
\end{figure}
\begin{table*}[t]
  \centering
  \caption{Speaker verification performance of the systems with different PTMs and input feature combinations with AS-norm.}
  \label{tab:main_results}
  \resizebox{\textwidth}{!}{
\renewcommand{\arraystretch}{1.07}
\small
\begin{tabular}{@{}c c c *{14}{c}@{}}
\Xhline{3.5\arrayrulewidth}
\multirow[c]{2}{*}{\raisebox{-0.8ex}{\centering Evaluation Set}} &
\multirow[c]{2}{*}{\raisebox{-0.8ex}{\centering PTM}} &
\multirow[c]{2}{*}{\raisebox{-0.8ex}{\centering System}} &
\multicolumn{7}{c}{EER (\%)} &
\multicolumn{7}{c}{MinDCF} \\
\cmidrule(lr){4-10}\cmidrule(lr){11-17}
& & & full & 5s & 4s & 3s & 2s & 1.5s & 1s & full & 5s & 4s & 3s & 2s & 1.5s & 1s \\
\midrule
\multirow{12}{*}{VoxCeleb-O}
& \multirow{4}{*}{wav2vec 2.0}
  & S1 & 1.37 & 1.30 & 1.38 & 1.46 & 1.83 & 2.20 & 3.33 & 0.1009 & 0.1037 & 0.1043 & 0.1097 & 0.1309 & 0.1529 & 0.2277 \\
& & S2 & 1.20 & 1.23 & 1.24 & 1.39 & \textbf{1.61} & 1.95 & 2.93 & 0.0863 & 0.0857 & 0.0915 & 0.0913 & 0.1150 & 0.1338 & \textbf{0.1942} \\
& & S3 & 1.31 & 1.35 & 1.38 & 1.48 & 1.84 & 2.23 & 3.28 & 0.0941 & 0.0915 & 0.0950 & 0.1066 & 0.1232 & 0.1405 & 0.2184 \\
& & S4 & \textbf{1.10} & \textbf{1.14} & \textbf{1.14} & \textbf{1.27} & \textbf{1.61} & \textbf{1.91} & \textbf{2.75}
& \textbf{0.0840} & \textbf{0.0846} & \textbf{0.0820} & \textbf{0.0900} & \textbf{0.1034} & \textbf{0.1323} & 0.1974 \\
\cmidrule(lr){2-17}

& \multirow{4}{*}{HuBERT}
  & S1 & 1.34 & 1.34 & 1.35 & 1.42 & 1.70 & 2.74 & 7.35 & 0.1007 & 0.0890 & 0.0871 & 0.0950 & 0.1261 & 0.1830 & 0.4476 \\
& & S2 & 1.32 & 1.25 & 1.22 & 1.36 & 1.65 & 2.57 & 5.89 & \textbf{0.0875} & \textbf{0.0833} & \textbf{0.0820} & \textbf{0.0923} & 0.1245 & 0.1930 & 0.3943 \\
& & S3 & 1.36 & 1.32 & 1.33 & 1.44 & 1.74 & 2.65 & 6.50 & 0.1017 & 0.0960 & 0.0970 & 0.1021 & 0.1276 & 0.1908 & 0.4102 \\
& & S4 & \textbf{1.18} & \textbf{1.14} & \textbf{1.20} & \textbf{1.27} & \textbf{1.55} & \textbf{2.23} & \textbf{4.60}
            & 0.0892 & 0.0903 & 0.0927 & 0.0988 & \textbf{0.1105} & \textbf{0.1580} & \textbf{0.2978} \\
\cmidrule(lr){2-17}
            
& \multirow{4}{*}{WavLM}
  & S1 & 0.97 & 0.96 & 0.98 & 1.05 & 1.36 & 1.60 & 2.49 & 0.0639 & 0.0659 & 0.0677 & 0.0829 & 0.0980 & 0.1219 & 0.1891 \\
& & S2 & 0.81 & 0.85 & 0.87 & 0.97 & 1.27 & 1.55 & 2.30 & 0.0614 & 0.0655 & 0.0687 & 0.0792 & 0.0954 & 0.1122 & 0.1692 \\
& & S3 & 0.85 & 0.89 & 0.93 & 1.01 & 1.28 & 1.51 & 2.31 & 0.0606 & 0.0651 & 0.0703 & 0.0756 & 0.0962 & 0.1185 & 0.1761 \\
& & S4 & \textbf{0.77} & \textbf{0.76} & \textbf{0.80} & \textbf{0.89} & \textbf{1.21} & \textbf{1.40} & \textbf{2.24}
& \textbf{0.0570} & \textbf{0.0556} & \textbf{0.0601} & \textbf{0.0644} & \textbf{0.0843} & \textbf{0.1093} & \textbf{0.1590} \\
\hline\hline

\multirow{12}{*}{VoxCeleb-E}
& \multirow{4}{*}{wav2vec 2.0}
  & S1 & 1.39 & 1.39 & 1.42 & 1.53 & 1.81 & 2.15 & 3.15 & 0.0881 & 0.0881 & 0.0907 & 0.0976 & 0.1194 & 0.1428 & 0.2003 \\
& & S2 & 1.30 & 1.30 & \textbf{1.33} & \textbf{1.43} & \textbf{1.66} & \textbf{1.96} & 2.82 & 0.0819 & \textbf{0.0810} & 0.0842 & \textbf{0.0909} & 0.1092 & 0.1285 & 0.1842 \\
& & S3 & 1.46 & 1.47 & 1.50 & 1.63 & 1.92 & 2.29 & 3.29 & 0.0900 & 0.0899 & 0.0925 & 0.1008 & 0.1213 & 0.1448 & 0.2059 \\
& & S4 & \textbf{1.28} & \textbf{1.29} & \textbf{1.33} & 1.44 & 1.68 & 1.97 & \textbf{2.80}
& \textbf{0.0814} & 0.0836\textbf{} & \textbf{0.0841} & 0.0923 & \textbf{0.1080} & \textbf{0.1276} & \textbf{0.1777} \\
\cmidrule(lr){2-17}
            
& \multirow{4}{*}{HuBERT}
  & S1 & 1.76 & 1.68 & 1.67 & 1.77 & 2.04 & 3.22 & 9.11 & 0.1160 & 0.1045 & 0.1043 & 0.1138 & 0.1350 & 0.2073 & 0.4872 \\
& & S2 & 1.28 & 1.23 & 1.25 & 1.33 & 1.58 & 2.26 & 5.35 & 0.0839 & 0.0803 & 0.0813 & 0.0874 & 0.1041 & 0.1503 & 0.3254 \\
& & S3 & 1.34 & 1.30 & 1.32 & 1.41 & 1.67 & 2.47 & 6.35 & 0.0870 & 0.0843 & 0.0856 & 0.0922 & 0.1098 & 0.1596 & 0.3646 \\
& & S4 & \textbf{1.19} & \textbf{1.18} & \textbf{1.20} & \textbf{1.29} & \textbf{1.52} & \textbf{2.12} & \textbf{4.30} 
& \textbf{0.0748} & \textbf{0.0743} & \textbf{0.0767} & \textbf{0.0831} & \textbf{0.0983} & \textbf{0.1410} & \textbf{0.2736} \\
\cmidrule(lr){2-17}
            
& \multirow{4}{*}{WavLM}
  & S1 & 1.05 & 1.05 & 1.07 & 1.16 & 1.37 & 1.65 & 2.44 & 0.0672 & 0.0664 & 0.0678 & 0.0735 & 0.0906 & 0.1103 & 0.1612 \\
& & S2 & 1.03 & 1.02 & 1.05 & 1.12 & 1.33 & 1.58 & 2.33 & 0.0640 & 0.0664 & 0.0661 & 0.0727 & 0.0872 & 0.1060 & 0.1548 \\
& & S3 & 0.98 & 0.99 & 1.01 & 1.09 & 1.29 & 1.56 & 2.37 & 0.0632 & 0.0628 & 0.0640 & 0.0692 & 0.0859 & 0.1044 & 0.1568 \\
& & S4 & \textbf{0.96} & \textbf{0.96} & \textbf{0.98} & \textbf{1.05} & \textbf{1.24} & \textbf{1.50} & \textbf{2.18}
& \textbf{0.0599} & \textbf{0.0606} & \textbf{0.0622} & \textbf{0.0672} & \textbf{0.0819} & \textbf{0.0973} & \textbf{0.1450} \\
\hline\hline

\multirow{12}{*}{VoxCeleb-H}
& \multirow{4}{*}{wav2vec 2.0}
& S1 & 2.55 & 2.56 & 2.62 & 2.82 & 3.30 & 3.89 & 5.47 & 0.1445 & 0.1486 & 0.1517 & 0.1641 & 0.1965 & 0.2287 & 0.3137 \\
& & S2 & 2.37 & 2.37 & 2.41 & 2.59 & 3.04 & 3.53 & 4.89 & 0.1384 & \textbf{0.1381} & 0.1416 & 0.1554 & 0.1835 & 0.2129 & 0.2840 \\
& & S3 & 2.62 & 2.66 & 2.72 & 2.93 & 3.43 & 4.01 & 5.50 & 0.1460 & 0.1485 & 0.1544 & 0.1655 & 0.1975 & 0.2303 & 0.3112 \\
& & S4 & \textbf{2.31} & \textbf{2.33} & \textbf{2.39} & \textbf{2.58} & \textbf{3.02} & \textbf{3.49} & \textbf{4.79}
& \textbf{0.1341} & 0.1396 & \textbf{0.1409} & \textbf{0.1523} & \textbf{0.1776} & \textbf{0.2064} & \textbf{0.2768} \\
\cmidrule(lr){2-17}
            
& \multirow{4}{*}{HuBERT}
  & S1 & 3.17 & 3.00 & 2.98 & 3.16 & 3.67 & 5.45 & 12.55 & 0.1817 & 0.1737 & 0.1747 & 0.1873 & 0.2182 & 0.3125 & 0.6222 \\
& & S2 & 2.33 & 2.25 & 2.26 & 2.40 & 2.87 & 3.99 & 8.00 & 0.1387 & 0.1340 & 0.1348 & 0.1458 & 0.1740 & 0.2407 & 0.4382 \\
& & S3 & 2.41 & 2.35 & 2.37 & 2.51 & 2.96 & 4.31 & 9.28 & 0.1415 & 0.1387 & 0.1412 & 0.1506 & 0.1785 & 0.2530 & 0.4896 \\
& & S4 & \textbf{2.18} & \textbf{2.16} & \textbf{2.21} & \textbf{2.36} & \textbf{2.77} & \textbf{3.80} & \textbf{6.79}
& \textbf{0.1263} & \textbf{0.1268} & \textbf{0.1297} & \textbf{0.1393} & \textbf{0.1663} & \textbf{0.2274} & \textbf{0.3884} \\
\cmidrule(lr){2-17}
            
& \multirow{4}{*}{WavLM}
  & S1 & 2.10 & 2.09 & 2.12 & 2.28 & 2.68 & 3.17 & 4.46 & 0.1271 & 0.1280 & 0.1309 & 0.1430 & 0.1688 & 0.1994 & 0.2718 \\
& & S2 & 2.01 & 2.01 & 2.05 & 2.22 & 2.60 & 3.02 & 4.30 & 0.1175 & 0.1174 & 0.1207 & 0.1311 & 0.1609 & 0.1867 & 0.2570 \\
& & S3 & 1.97 & 2.00 & 2.05 & 2.20 & 2.57 & 3.03 & 4.35 & 0.1171 & 0.1198 & 0.1215 & 0.1317 & 0.1579 & 0.1866 & 0.2590 \\
& & S4 & \textbf{1.89} & \textbf{1.89} & \textbf{1.92} & \textbf{2.08} & \textbf{2.43} & \textbf{2.86} & \textbf{4.02}
& \textbf{0.1115} & \textbf{0.1129} & \textbf{0.1151} & \textbf{0.1253} & \textbf{0.1496} & \textbf{0.1736} & \textbf{0.2417} \\
\Xhline{3.5\arrayrulewidth}
\end{tabular}
}
\end{table*}

\section{Methods}
\label{sec:format}
Utilization of the MRE along with the FBank features was shown to be effective for short-segment SV \cite{ecapa-mre}, but the performance may be further improved if a large scale speech data is exploited through PTMs. 
The integrated features combining FBank and PTM features in \cite{ssl-sv4} exhibited wonderful performance, but the temporal resolution was limited to 10 ms and 20 ms which may not be ideal for short-segment SV. 
In this paper, we propose to utilize PTM features and FBank features along with an MRE to extract as much information as possible from short utterances for SV.
As a backbone SV system, we have used ECAPA-TDNN, although any other SV system can be tested.
The block diagram of the proposed system is illustrated in Fig. \ref{fig:overall_framework}, which can be simplified to Fig. \ref{fig:res_d} emphasizing combination of input features. Fig. \ref{fig:res} also shows other input feature combinations to utilize PTM features.

The input speech signal $\mathbf{x}$ is first fed into a log mel-filterbank feature extractor, an MRE module, and a PTM to extract FBank feature $\mathbf{F}$, multi-resolution features $\mathbf{Z}_\text{cond}$, and layer-wise speech representations  $H_0, \ldots, H_L$. 
As for PTM features, $\mathbf{x}$ is processed by a CNN encoder and $L$ Transformer layers of a PTM to produce $L+1$ layer-wise speech representations $H_0, \ldots, H_L\in \mathbb{R}^{D \times T }$, where $D$ and $T$ denote the channel and the number of frames, respectively.  
The speech representations are 
then fused by a weighted summation with learnable parameters $\mathbf w=\left[w_0, \ldots, w_L\right]$ to form a PTM feature matrix $\mathbf{S}$, i.e., 
\begin{equation}
\mathbf{S} = \sum_{l=0}^{L} {w}_l {H}_{l} \in\mathbb{R}^{D \times T }
\end{equation}
As the FBank feature $\mathbf{F}\in \mathbb{R}^{C_\text{mel}\times 2T}$ has $C_\text{mel}$ channels and twice the number of frames compared with $\mathbf{S}$ with 10 ms frame shift, the temporal and channel dimensions are adjusted using a CNN extractor as in \cite{ssl-sv4} to form $\mathbf{F'}\in \mathbb{R}^{D\times T}$. $\mathbf{S}$ and $\mathbf{F'}$ are added up to form the input of the SV system without weights expecting the CNN extractor would control the relative weights for two features. 

The MRE output $\textbf{Z}_\text{cond} \in\mathbb{R}^{NQ \times T'}$ is downsampled to match the temporal dimension $T$ to form $\textbf{Z}_\text{cond}' \in\mathbb{R}^{NQ \times T}$.  
 $\textbf{Z}_\text{cond}'$ is then used to modify the hidden representations before each SE-Res2Block in the ECAPA-TDNN through adapter modules \cite{ecapa-mre}, i.e., to modify the hidden representation $\mathbf{h}\in \mathbb{R}^{C \times T}$ with an affine transformation 
\begin{equation}
\mathbf{\bar{h}} = \gamma\otimes\mathbf{h}+\beta
\end{equation}
where $\otimes$ means the element-wise multiplication and $\gamma\in \mathbb{R}^{C \times T}$ and $\beta\in \mathbb{R}^{C \times T}$ are obtained from $\textbf{Z}_\text{cond}'$.
The final output of the ECAPA-TDNN, $\mathbf{e}$, is used for SV. 
\section{Experimental Settings}
\label{sec:pagestyle}

All systems were trained on the VoxCeleb2 development set \cite{vox2}, comprising more than 1 million utterances from 5,994 speakers.
For evaluation, we employed the VoxCeleb1-O, VoxCeleb1-E, and VoxCeleb1-H test sets, which consist of 37,611, 579,818, and 550,894 verification trials spoken by 40, 1,251, and 1,251 speakers, respectively.
There is no speaker overlap between the training and test sets.
The SV decision is made by thresholding the cosine similarity between the embeddings from the test and enrollment utterances after optionally applying the adaptive s-norm (AS-norm) \cite{as-norm}, \cite{s-norm}. 
The performance on short-segment SV was evaluated by averaging the performances for the experiments with full-length enrollment utterances and short test utterances cropped in the middle, and vice versa.
The equal error rate (EER) and the minimum of the detection cost function (MinDCF) with $P_\text{target} = 0.05$ and $C_\text{FalseAlarm} = C_\text{Miss} = 1$ were used as performance measures.

Three large PTMs, WavLM-Large \cite{wavlm}, HuBERT Large \cite{HuBERT}, and wav2vec 2.0 Large \cite{w2v2} models\footnote{\href{https://huggingface.co/models}{huggingface.co/models}} were adopted in the experiments.
The parameters for PTMs were not updated. 
ECAPA-TDNN with $C=512$ channels was used as the backbone SV model. All hyperparameters for the MRE module were set identically to those for the optimal configuration reported in \cite{ecapa-mre}, in which 4 SREs with the window shifts of 25, 50, 100, and 200 samples were used, which correspond to the temporal resolutions of 1.56, 3.13, 6.25, and 12.5 ms for the 16 kHz sampling rate, respectively.
Each utterance was randomly cropped into 2-second segments to construct training batches, i.e., $ L_x = 32,000$ samples in training. 
For FBank features, 80-dimensional log mel-filterbank energies were extracted using a 25 ms window and a 10 ms frame shift.
We applied data augmentation using simulated room impulse responses (RIRs), MUSAN noises \cite{musan}, and speed perturbation with factors of 0.9 and 1.1. 
Voice activity detection (VAD) was not applied. 
The models were optimized using AdamW with a cyclical learning rate policy \cite{clr}, where the maximum and minimum learning rates were set to $1\times10^{-3}$ and $1\times10^{-8}$, respectively, and the maximum learning rate was reduced by half every cycle. 
We adopted AAM-Softmax \cite{aam-softmax} as the loss function with a margin of 0.2 and a scale of 30. 
Training was conducted for six cycles of 5 epochs each, totaling 30 epochs.
All systems were implemented in PyTorch framework and trained on two NVIDIA RTX 3090 GPUs.

\section{Results and Analysis}
\label{sec:majhead}
\label{ssec:subhead}
Table \ref{tab:main_results} presents the speaker verification performances with three PTMs and four input feature combinations shown in Fig. \ref{fig:res} using AS-norm for various input lengths. 
As can be seen in the Table, the proposed system, S4, showed the best performance for most of the cases when the PTM was wav2vec 2.0 or HuBERT and all cases when the PTM was WavLM. Among three PTMs, WavLM exhibited the uniformly best performance. 
Introducing MRE features lead to performance improvement in most of the cases, which can be observed from the performance improvement of S2 over S1 and that of S4 over S3. 
On the other hand, the introduction of FBank features (S3 compared with S1, and S4 compared with S2) was not very effective for wav2vec 2.0, although it improved the SV performance for WavLM. 

The WavLM-based system with four input feature combinations was also compared against previously proposed short-segment SV systems for the VoxCeleb1-O dataset, as shown in Table \ref{tab:voxceleb1o}.
The reported performances were obtained from the original papers \cite{rawnext, ecapa-mre, mr-rawnet}, except for ECAPA-TDNN, where we used the re-implemented results in \cite{ecapa-mre}.
AS-norm was not applied in this experiment, as the compared systems did not utilize it. 
The ``Utilized Features" columns are to show that the ECAPA-TDNN \cite{ecapa-tdnn} and ECAPA-TDNN + MRE \cite{ecapa-mre} can also be though as a different input feature combination for the same system, although the ECAPA-TDNN in \cite{ecapa-tdnn} and \cite{ecapa-mre} was with $C=1024$ channels instead of $512$ channels. These columns are not meaningful for RawNext \cite{rawnext} or MR-RawNet \cite{mr-rawnet}, but the MRE column is checked for MR-RawNet \cite{mr-rawnet} to show that it uses the MRFE which is similar to the MRE. 
The proposed S4 system consistently outperformed all other systems for various input lengths including ECAPA-TDNN + MRE \cite{ecapa-mre}, which previously showed the best performance for the input length of 2 s or shorter. With the PTM features alone, S1 did not outperform ECAPA-TDNN + MRE \cite{ecapa-mre} for short-segment SV. 

\begin{figure}[t]
    \centering
    \includegraphics[width=\columnwidth]{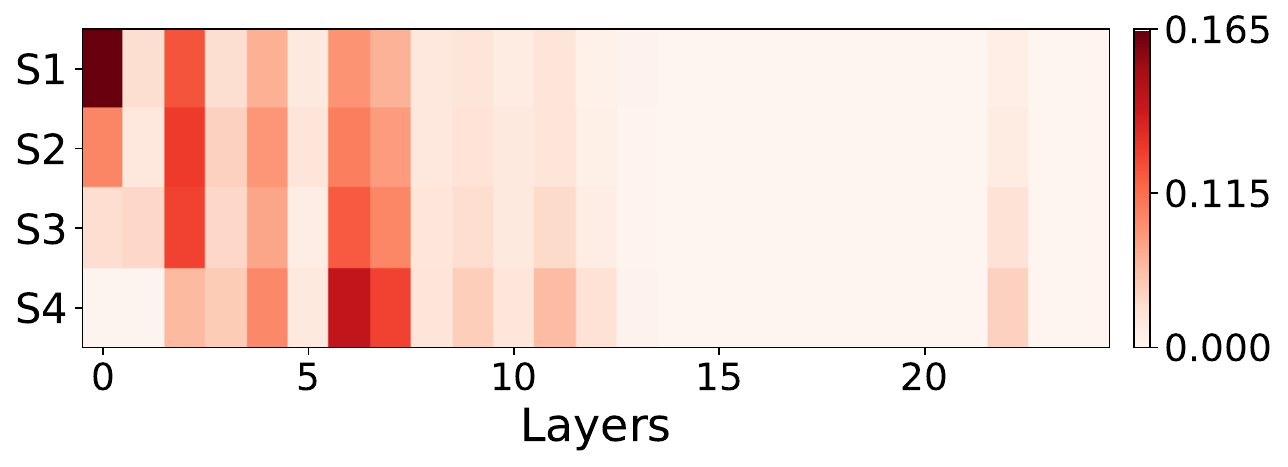}
    \vspace{-0.5cm}
    \caption{Learned layer-wise weights for WavLM-Large features in the systems with different combinations of input features. }
    \label{fig:WavLM-weight}
\end{figure}

\label{ssec:subhead}
To investigate the effect of the additional features to the PTM-based SV system, we plot the learned layer-wise weights, $\mathbf w=\left[w_0, \ldots, w_L\right]$, for four input combinations when the PTM was the WavLM in Fig. \ref{fig:WavLM-weight}. 
The learned weights for S1 showed that $w_0$ had the biggest value, which was similar to the weights reported in the original WavLM paper\cite{wavlm}. 
With the additional MRE or FBank features, S2 and S3 emphasized higher level informations more than $H_0$. With both MRE and FBank features, the layer-wise weights for the proposed S4 system $H_0$ was barely used, which may imply that the low-level information was exploited enough through the MRE and FBank features. Instead, higher level information can be more emphasized in $\mathbf{S}$ to complement $\mathbf{F}$ and $\mathbf{Z}_{cond}$. 

\begin{table}[t]
  \large
  \centering
  \caption{Performance comparison with previous approaches for short segment speaker verification on VoxCeleb1-O without AS-norm. $^{\dag}$: re-implementation.}
  \label{tab:voxceleb1o}
  \vspace{-0.25cm}
  \resizebox{\linewidth}{!}{%
    
\renewcommand{\arraystretch}{1.2}
\begin{tabular}{c c c c c c c c c}
\Xhline{3.5\arrayrulewidth}
\multirow{2}{*}{\raisebox{-0.3ex}{\centering System}} & \multicolumn{3}{c}{\raisebox{-0.3ex}{Utilized Features}} & \multicolumn{5}{c}{\raisebox{-0.3ex}{EER (\%)}} \\
\cline{2-9}
 & PTM & FBank & MRE & full & 5s & 2s & 1.5s & 1s \\
\midrule
\multicolumn{1}{r}{RawNext \cite{rawnext}} & \textendash & \textendash & \textendash & 1.29 & 1.45 & 2.34 & \textendash & 4.47 \\
\multicolumn{1}{r}{MR-RawNet \cite{mr-rawnet}} & \textendash & \textendash & \cmark & 0.83 & 0.99 & 1.61 & \textendash & 3.47 \\
\hline
\multicolumn{1}{r}{ECAPA-TDNN$^{\dag}$\cite{ecapa-tdnn}} & \xmark & \cmark & \xmark & 1.03 & 1.05 & 1.76 & 2.12 & 3.04 \\
\multicolumn{1}{r}{+ MRE \cite{ecapa-mre}}      & \xmark & \cmark & \cmark & 1.01 & 1.03 & 1.32 & 1.60 & 2.33 \\
\midrule
S1 & \cmark & \xmark & \xmark & 0.98 & 1.00 & 1.40 & 1.65 & 2.52 \\
S2 & \cmark & \xmark & \cmark & 0.88 & 0.91 & 1.30 & 1.56 & 2.36 \\
S3 & \cmark & \cmark & \xmark & 0.89 & 0.92 & 1.32 & 1.54 & 2.38 \\
S4 & \cmark & \cmark & \cmark & \textbf{0.78} & \textbf{0.81} & \textbf{1.24} & \textbf{1.46} & \textbf{2.29} \\
\Xhline{3.5\arrayrulewidth}
\end{tabular}

  }
\end{table}
\section{Conclusion}
\label{sec:print}
In this paper, we have introduced a short-segment SV system which utilizes the self-supervised learning models trained with large-scale databases and supplements their shortcoming for the short-segment SV, insufficient temporal resolution, by utilizing log mel-filterbank features and a multi-resolution encoder with various kernel sizes. 
The features from the PTMs are combined with the FBank features to form the input tensor for the backbone SV model, while the features generated by the MRE are used to modify the hidden representations within the SV model. 
Experimental results on the VoxCeleb dataset with WavLM for various input lengths demonstrated that the proposed system outperformed previously proposed methods on short-segment SV. 
Furthermore, an analysis of the learned weights to combine layer-wise outputs of the WavLM for
different combinations of the input features revealed that the inclusion of the MRE and FBank features lead PTM representations to focus more on the high-level information.



\vfill\pagebreak




\bibliographystyle{IEEEtran}
\bibliography{strings,refs}

\end{document}